\documentclass[aps,twocolumn,nopacs,floats,superscriptaddress]{revtex4-1}
\usepackage{graphicx}
\usepackage{dcolumn}
\usepackage{bm}
\usepackage{color}
\def\he4{$^4$He}
\def\hel3{$^3$He}
\def\Am3{\AA$^{-3}$}
\def\beq{\begin{equation}}
\def\eeq{\end{equation}}
\begin{document}

\author{Anatoly B. Kuklov}
\affiliation{Department of Physics \& Astronomy, College of Staten Island and the Graduate Center of
CUNY, Staten Island, NY 10314}

\author{Nikolay V. Prokof'ev}
\affiliation{Department of Physics, University of Massachusetts, Amherst, MA 01003, USA}
\affiliation{National Research Center ``Kurchatov Institute," 123182 Moscow, Russia}
\author{Boris V. Svistunov}

\affiliation{Department of Physics, University of Massachusetts, Amherst, MA 01003, USA}
\affiliation{National Research Center ``Kurchatov Institute," 123182 Moscow, Russia}
\affiliation{Wilczek Quantum Center, School of Physics and Astronomy and T. D. Lee Institute, Shanghai Jiao Tong University, Shanghai 200240, China}


\title{Disorder-induced quantum properties of solid \he4}

\begin{abstract}
We briefly discuss theoretical and experimental discoveries in the field of supertransport and plasticity in imperfect solid \he4
and argue that these promise new exciting  developments. Several experiments aimed at clarifying the origin of the
supertransport and its relation to plasticity are proposed. In particular, we argue that  ``cold-working" protocols
of sample preparation should be crucial in this respect. \\
\end{abstract}

\maketitle
\section{Introduction}
The idea of supersolid as a stable phase of a crystal supporting supertransport of its own atoms was proposed more than
60 years ago by E. Gross \cite{Gross} and further elaborated microscopically in Refs.~\cite{Andreev, Thouless, Chester}.
Early attempts to detect such a phase in solid \he4 yielded zero result.
The interest to the subject exploded in 2004, after Kim and Chan claimed the observation of the supersolid response
in the torsional oscillator experiment \cite{KC}. Their work initiated an intensive experimental and theoretical
activity worldwide. It was quickly established through rigorous theoretical considerations and large-scale {\it ab initio} simulations \cite{Cep,PS2005,SG,Clark,fate} (see also Ref.~\cite{Kolya}) that ideal crystals of \he4 are not supersolids,
and only disorder could induce the superfluid response.

Superfluidity was numerically observed in the maximally disordered simulation sample (termed ``superglass") \cite{SG},
as well as along some, but not all, grain boundaries \cite{GB}.
Subsequent numeric studies revealed superfluidity along the core of the screw dislocation with the Burgers
vector oriented along the {\it hcp} axis \cite{screw}. 

To explain these findings, a generic criterion for inducing superfluidity in solid \he4 has been established
\cite{stress}---a structural defect producing strain larger than the 10-15\% threshold results in closing the local
vacancy gap in the vicinity of the defect. In particular, this explained the insulating character of basal edge
dislocations, which split into partials and, thus, produce weaker strain in the vicinity of its partial cores.

On the experimental front, a dc superflow through the solid \he4 has been discovered by Ray and Hallock \cite{Hallock}
in the unique ``UMass sandwich" setup that uses Vycor  ``electrodes" inserted into the solid sample.
The resulting flux of atoms through the solid was extremely weak---about few ng/s, and the effect was not reproducible
from sample to sample; certain samples demonstrating no effect at all. This was the strongest indication that
disorder was the culprit. (Consistent with that, the claim of the supersolid phase of \he4 as a cause of the
torsional oscillator anomaly has been eventually retracted \cite{KC2012}).

\begin{figure}[!htb]
	\includegraphics[width=1.0 \columnwidth]{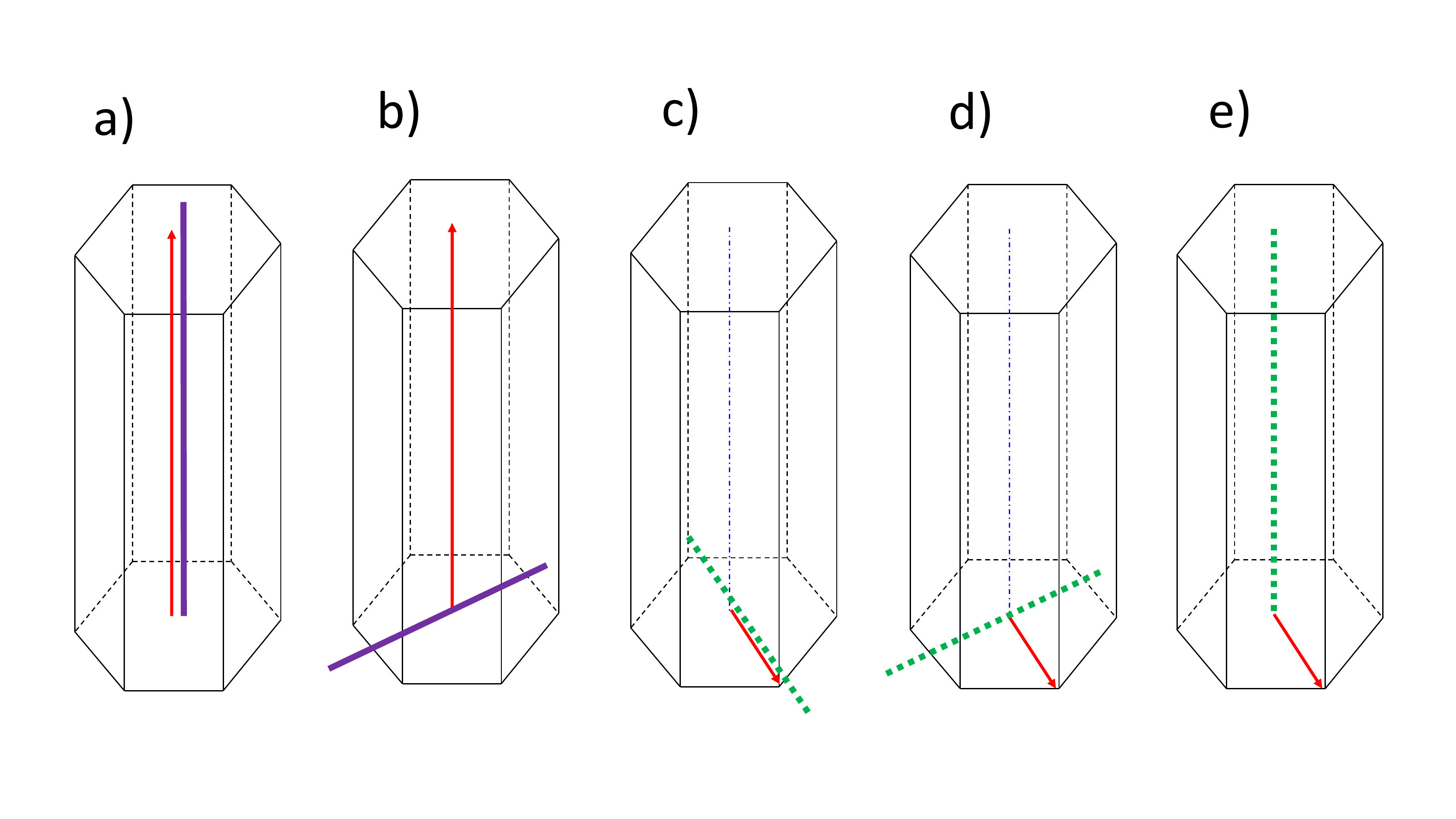}
	\vskip-8mm
	\caption{ (Color online) Typical dislocations (thick lines) and their Burgers vectors (solid arrows) in {\it hcp} solid \he4.  The panels (a) and (b) depict dislocations with superfluid cores (thick solid lines)---the screw \cite{screw} and superclimbing edge \cite{sclimb}, respectively. 
The dislocations with insulating cores (thick dotted lines) are shown in panels (c), (d), and (e)---the basal screw, basal edge, and non-basal edge, respectively.}
\label{disl}
\end{figure}

Among a number of unusual supertransport features discovered in the experiment \cite{Hallock}, the
uniform matter accumulation in solid \he4 is the most surprising.  This anomalous (or giant) isochoric
compressibility has been dubbed the {\it syringe}  effect in Ref.~\cite{sclimb}, where it has been proposed
and corroborated by first-principle simulations that the effect is due to the superclimb of edge dislocations, which have superfluid core.
[As an illustration, Fig.~\ref{disl} shows typical dislocations and indicates
their properties with respect to superflow].

As opposed to the conventional climb mechanism supported by pipe diffusion of thermally activated vacancies and atoms
along the dislocation cores \cite{Lothe,Hull}, the superclimb is a novel low-temperature phenomenon
supported by the superfluid transport along the dislocation cores, and as such is a unique property of solid \he4.
It is important to note that the syringe effect is responsible for crystal growth above the melting line
in the experimental protocols of Refs.~\cite{Hallock,Hallock2010}.
In its inverse form, the syringe effect---along with its unusual temperature dependence \cite{Hallock}---has been
confirmed by Beamish, Balibar, and their collaborators in a completely different setup \cite{Beamish}.
As temperature decreases, the superclimb rate increases \cite{Hallock,Beamish}, which is very different from
the conventional climb rate vanishing according to the activation law \cite{Lothe,Hull}.

Recently, Chan and collaborators studied superflow properties of solid samples of different length \cite{Moses,Moses2019},
and the main results of Refs.~\cite{Hallock, Beamish} have been confirmed.  At this juncture, it is useful to outline some
quantitative characteristics observed in the experiments \cite{Hallock,Beamish,Moses,Moses2019}. The temperature dependence of the superflow rate, $F(T)$, shows an almost linear suppression as the temperature $T$ increases, with the flow vanishing at about $T=T_1\approx 0.6  \div 1.0$K, depending on the sample length. This can be roughly described as $F \sim 1-T/T_1$ for $T<T_1$.  The flow is proportional to a sublinear power of the chemical potential bias, $\mu$, between the Vycor rods: $F \sim \mu^b$ with $b<0.5$ \cite{Hallock,Moses,Moses2019}.
If taken apart from the temperature dependence, this feature is consistent with the supercritical flow through
Luttinger liquid in the regime where the source of phase slips is a single weak link \cite{Kane,Kash,Kolya2000}.
However, the conventional model of such a regime \cite{Kane,Kash,Kolya2000} implies no variation of the flow rate with temperature,
while the experimental data in \he4 \cite{Hallock, Moses,Moses2019} can be represented as a
simple product of the $T$ and $\mu$ factors:
\beq
F(T,\mu) \sim (1-T/T_1) \mu^b.
\label{F}
\eeq
The origin of (\ref{F}) remains unclear. The only general statement that one can make at this point is that if
the factor $\mu^b$ is indeed due to the phase slips in the Luttinger liquid, then the origin of the factor $(1-T/T_1)$ should be (quasi-)extrinsic.

The flow rate depends strongly on the pressure in the solid, and with good accuracy it can be described by the exponential suppression \cite{Moses} of the overall factor in Eq.(\ref{F}). The temperature $T_1$ in Eq.(\ref{F}) shows a significant decrease with pressure too. It has also been found that the flow rate demonstrates log-suppression with the length of the conducting pathways \cite{Moses2019}. The analysis, however, was conducted using data collected from two different groups of samples (showing large variability of the flow rates).

The key question to answer is that of the structure of the conducting channels responsible for the supertransport.
An appealing model of percolating multiply-connected network formed by edge dislocations with superfluid cores
was proposed by Shevchenko more than 30 years ago \cite{Shevchenko_87}.  If a typical size $R$ of the network segments is much larger than an interatomic distance $a$, the transition temperature $T_c$ below which the network develops a global coherence is suppressed relative to the
characteristic microscopic \he4 temperature $T_0\sim 1$K as $T_c\approx T_0 \, a/R$. However, in a wide temperature range $T_c<T \ll T_0$, transport properties are characterized by long relaxation times controlled by phase slips \cite{Shevchenko_88,Fil}.
The random dislocation network itself is considered to be static in this model.

What is missing in the Shevchenko scenario is the superclimb \cite{sclimb} of the edge segments, which, as will be detailed below,
can lead to the destabilization of the network through the tendency to decrease the total length of the dislocations.
The required stability may be provided by non-superfluid dislocations forming its own stable network.
An alternative dynamic scenario considers superclimbing loops (prismatic loops in the basal plane) that are
injected into the solid through Vycor rods and inflated by the chemical potential bias.
Such loops then establish temporary flow pathways through proximity tunneling of \he4 atoms between the loops.

One of the fundamental questions in the field of solid \he4 currently is the relationship between the
supertransport phenomena and plasticity. Experiments on the nature of plasticity in solid \he4 (for overview, see Ref.~\cite{Beamish2019})
have revealed complex processes involving creation and recombination of dislocations as well as dislocation avalanches,
providing important information on spatial scales involved \cite{Beamish2018} -- covering at least three orders of magnitude from few $\mu$m to few mm. 

Here we suggest several experiments which should help establish the nature of the superfluid transport through solid \he4 and its connection with the plasticity.
One of them addresses the possibility that plastic deformation may induce the superfluid pathways.

\section{Geometry of the conducting dislocation network}
In clean hcp monocrystals at low $T$, gliding basal dislocations can form a stable network of Y-type junctions (see in Refs.~\cite{Lothe,Hull}). Such junctions are possible because the condition of the Burgers vector conservation can be met at each junction. However, basal dislocations have an insulating core (see in Fig.~\ref{disl}) and, thus, cannot be responsible for the superflow. The situation for the superclimbing dislocations is completely different. Their Burgers vector has only one orientation---along (or against) the hcp symmetry axis, as indicated in Fig.~\ref{disl}. In this case, no stable junctions can be formed.

\subsection{Instability of the network of superclimbing dislocations}
An intersection of superfluid screw and edge dislocations should result in creating a kink on the edge dislocation (see in Ref.~\cite{Hull}),
and this may produces a marginal junction between the two. A 3D network consisting of a forest of screw dislocations pinned by prismatic loops (made of superfluid edge dislocations) has been proposed in Ref.~\cite{Kuklov2011}. In this case, however, the loops located in different remote basal planes can exchange particles through the screw dislocations.  In addition, two prismatic loops approaching each other to a distance less than
several interatomic distances can exchange particles through the proximity  effect. Both channels for particle exchange lead to the destabilization of the  network with respect to merging neighboring segments and to reducing the network length and, accordingly, to enlarging length of individual segments. Eventually, only independent long dislocations disconnected from each other should remain in the crystal.

To illustrate the point, consider the network formed by the prismatic loops. They are
tight clusters of vacancies or interstitials in the basal plane, and their rims are partial edge dislocations with the (half) Burgers vector along the hcp axis. Thus, the rims must be superfluid and can perform superclimb.
In a perfect solid and at low temperature, such a loop can move ballistically by dissipationless matter transfer from one end
of the loop to the other. It is worth noting that prismatic loops can be viewed as dislocation dipoles (see Refs.~\cite{Landau,Lothe}).
Such dipoles interact by sign varying long-range forces.
In particular, loops belonging to the same basal plane and characterized by the same Burgers vector repel each other
at distances much larger than $R$. However, if two such loops are at a distance comparable or smaller than $R$,
the repulsion is changed to attraction and, eventually, they should merge together to form a larger loop with the total length
of the rim reduced from $2R$ to $\sqrt{2}R$.

While the lowest energy state of a dilute system of $N$ loops corresponds to one macroscopic loop with size $\propto \sqrt{N}$,
it is not easy to reach because it is separated by a large energy barrier.
The best metastable configuration is a dipole solid with inter-loop separation $D >R$.
The leading destabilizing mechanism is particle transfer between the loops capable of continuously
reducing the system energy. Indeed, the transfer of $\Delta N$ particles between two loops of size $R$ changes
the total rim energy by
\[
\Delta U \sim \sqrt{R^2 + a^2 \Delta N} + \sqrt{R^2 - a^2\Delta N} - 2R \approx - \frac{a^4 \Delta N^2}{4R^3}.
\]
Even in the absence of particle transfer within the percolating network, tunneling of individual atoms
between the loops makes the dipole solid unstable. However, the tunneling amplitude is exponentially small in $D$;
as the number of remaining loops decreases, the inter-loop separation increases, and the relaxation
time quickly exceeds the experimental time scale. Simultaneously, the flow through the sample should essentially vanish.

\subsection{Compound network of basal and superclimbing dislocations}
The above discussion did not take into account that segments of superclimbing dislocations can be trapped in potential
wells created during crystal growth. The component of the stress tensor $\sigma_{\mu \nu}$ responsible for the
trapping is $\sigma_{zz}$ where $Z$-direction is along the hcp axis (see Ref.~\cite{Lothe}).
The lines where $\sigma_{zz}$ changes sign trap the dislocation core and suppress the superclimb.
Without such traps, the threshold for external bias to initiate superclimb scales as $ \sim 1/R$, where $R$
is the length of a free segment \cite{Kuklov2015}. In contrast, freeing the dislocation from the linear trap
requires a much larger bias determined by the depth of the potential well and independent of the dislocation length.

Trapping of the superclimbing dislocations can be induced by basal (non superfluid) dislocations forming
a stable 2D network made of so called Y-junctions (see Refs.~\cite{Lothe,Hull}).
A pair of basal and superclimbing dislocations can be bound by elastic forces \cite{Kuklov2019}.
This process does not exclude the superclimb phenomenon because both dislocations can move in tandem in response
to the bias by the chemical potential. Thus, it is natural to anticipate that a stable network of basal dislocations
can stabilize a network of superclimbing dislocations. 

There are two options for such a network: (i) long superclimbing dislocations separated from each other are trapped by the basal network and establish the connection between the Vycor electrodes; (ii) mesoscopic prismatic loops are stabilized by the basal network, and the percolation of the flow is established due to Josephson effect between neighboring loops.

 In the first case the syringe effect is determined by a typical size $R$ of bound segments of superclimbing and basal dislocations. Such a coupling is  responsible for the transverse supershear effect \cite{Kuklov2019}. It is important that the superflow is determined essentially by properties of single dislocation (enhanced by their number in a sample). The weak link (or two) should be located at the boundaries between the Vycor electrodes and the solid.

\begin{figure}[!htb]
	\includegraphics[width=0.75 \columnwidth]{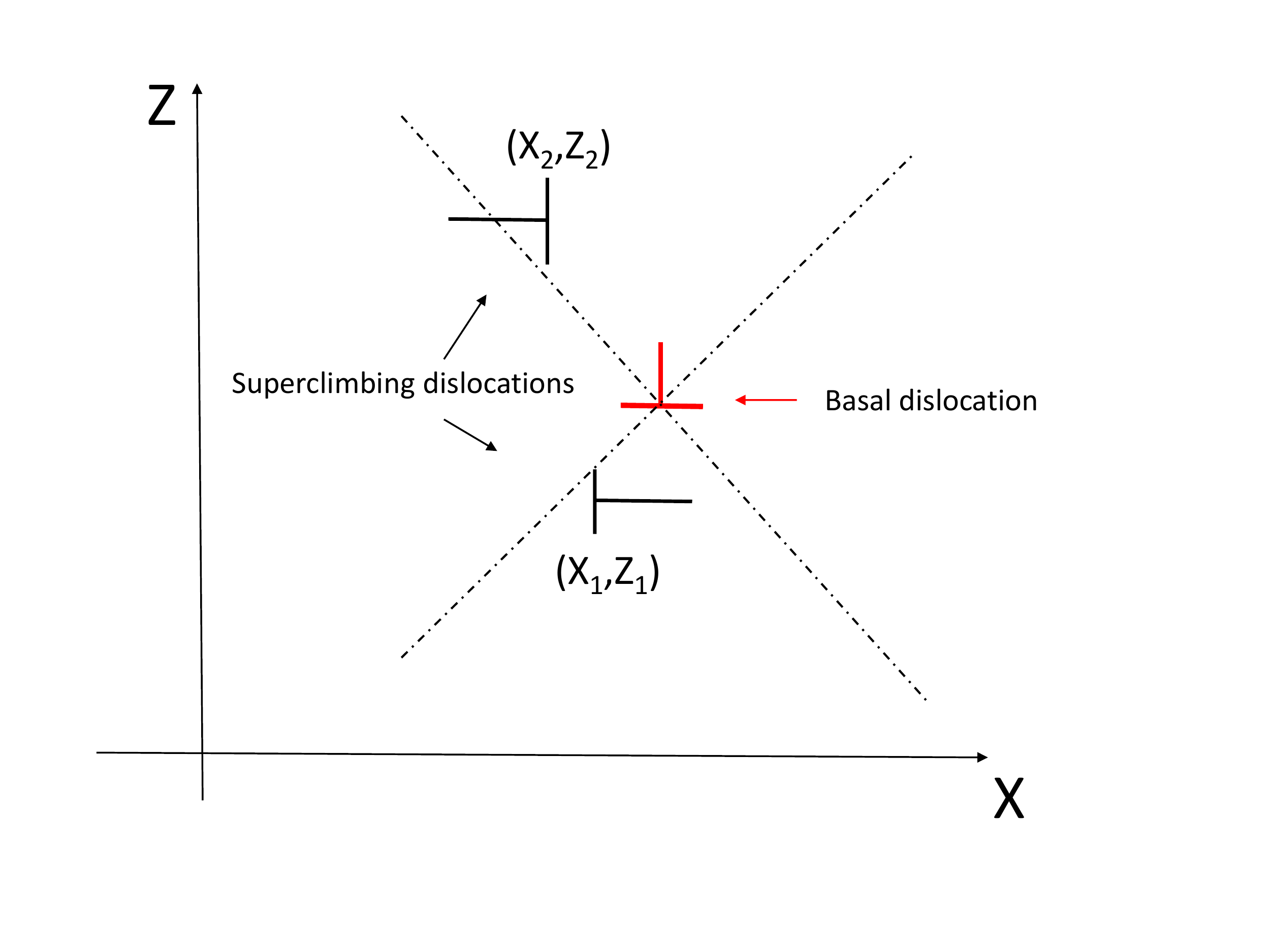}
	\vskip-8mm
	\caption{(Color online) Bound state of one basal and two superclimbing dislocations (shown by T-shape lines) with parallel cores. Dot-dashed lines, $|z|=|x|$ with the origin located at the basal dislocation, indicate equilibrium positions of superclimbing dislocations in the elastic field provided by the basal dislocation \cite{Kuklov2019} for two possible orientations of their Burgers vector
(ignoring the force between the superclimbing dislocations). For the dislocation located at $(x_2,z_2)$ ($(x_1,z_1)$)
stable equilibrium is close to the dot-dashed line with negative (positive) slope.
Attractive interaction between superclimbing dislocations shifts their equilibrium positions away from the dot-dashed lines
toward each other (with exception in the case $z_1=z_2$).  }
	\label{three}
\end{figure}

\begin{figure}[!htb]
\includegraphics[width=0.75 \columnwidth]{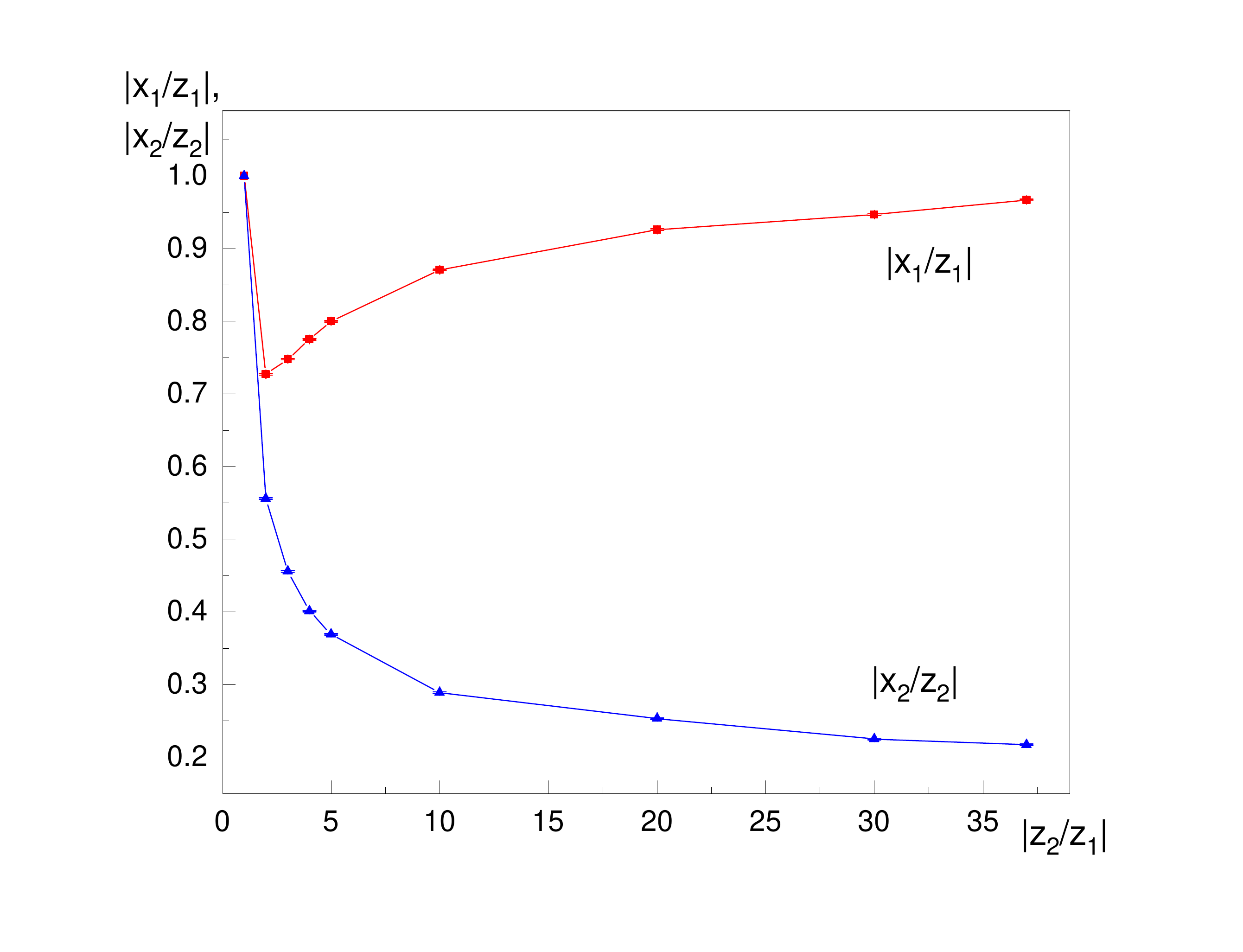}
\vskip-8mm
\caption{(Color online) Equilibrium positions of the superclimbing dislocations with respect to the basal dislocation
(as in Fig.~\ref{three}) expressed as ratios.
}
\label{ratios}
\end{figure}

The case (ii) is corroborated by extending the result of Ref.~\cite{Kuklov2019} to the case of one basal and two superclimbing dislocations. As it turns out there is a wide range of parameters where all three of them form a bound state. 
The arrangement shown in Fig.~\ref{three} can be analyzed within the isotropic medium approximation
if all distances are much larger than $a$. The stable equilibrium condition is found by using the solution for elastic
stress field produced by straight infinitely long dislocations that all can move along $X$-direction either by glide or superclimb.
In Fig.~\ref{three} the basal dislocation is located at $x=0, z=0$ and two superclimbing ones are at $(x_1,z_1)$, $(x_2,z_2)$, where without loss of generality it is assumed $|z_2|\geq |z_1|$.
The Z-axis in Fig.~\ref{three} is along the hcp axis of the solid. Glide of the superclimbing dislocation along Z-direction is strongly suppressed by the high Peierls barrier. [Furthermore, since the suprclimbing dislocation is split into partials \cite{sclimb} with the fault plane in between whole fault must be dragged along Z to support glide]. The key results are as follows (see also Ref.~\cite{Kuklov2019}).
No stable equilibrium exists if both superclimbing dislocations are located above or below the basal dislocation
unless $20<z_2/z_1<38$. In this case, the Josephson coupling between them can be safely neglected.
When superclimbing dislocations are on the opposite sides of the basal one (as in Fig.~\ref{three}) it is found that  stable  equilibrium exists for $1 \leq |z_2/z_1|\leq 37$. The equilibrium positions along X (in units of their Z-coordinates) of the superclimbing dislocations are shown in Fig.~\ref{ratios} as functions of $|z_2/z_1|$,  Thus, when all distances are of the order of $a$, the
Josephson coupling between the superclimbing dislocations may be substantial; otherwise it is exponentially suppressed.
Thus, such a mechanism may create a percolating multiconnected network of superfluid dislocations. However, it should
be realized that the superflow in this case will be dominated by the Josephson junctions between the trapped prismatic loops.
This raises a question how macroscopic number of such loops are formed close enough to each other. From our perspective such
a possibility represents the ``fine-tuning" scenario, and we consider it as highly unlikely to occur during random process of solid growth.

\section{Perspectives}
We find of utmost importance to design and conduct experiments revealing the nature of the conducting pathways
and the origin of the observed temperature and bias dependencies \cite{Hallock,Moses,Moses2019, Beamish}. Of equal importance is to
understand the interplay between the supertransport and low-temperature plasticity.
Several possible setups along these lines are outlined below.

\subsection{Superflow and syringe modulated by uniaxial stress}
To detect the presence of the compound network of the type ii) consider applying an external uniaxial stress $\sigma^{(e)}_{zz}$
(with all other  stress components being zero). As is clear from Fig.~\ref{three}, if $\sigma^{(e)}_{zz}>0$,
there will be an additional force on the pair of superclimbing dislocations trying to increase their separation along the basal plane (X-direction in Fig.~\ref{three}). Since the Josephson coupling is exponentially suppressed with the tunneling distance,
the superflow rate is expected to decrease sharply if such coupling is important for establishing the flow pathways.
If the sample is a monocrystal, imposing stress in other directions should not result in any effect. 
If the global connectivity is due to the forest of screw dislocations or long wall-to-wall superclimbing, then 
the flow will be rather insensitive to the $\sigma^{(e)}_{zz}$ stress. 

It is worth noting that the inverse syringe effect is a response of the edge dislocations (of the type b), Fig.\ref{disl}) on the stress $\sigma^{(e)}_{zz}$ changing chemical potential. Thus, observing the asymmetry of  the inverse syringe effect in a monocrystal would provide the direct evidence for the mechanism
behind the observation \cite{Beamish} made in the polycrystalline \he4.

\subsection{Transverse supershear effect}

Bound states of basal and superclimbing dislocations have unusual response to the shear stress by producing
the syringe effect. It is called the transverse supershear phenomenon in Ref.~\cite{Kuklov2019}.
This effect is based on binding between basal and superclimbing dislocations. Once the shear stress $\sigma^{(e)}_{zx}$ is 
imposed to move basal dislocations along X-axis, superclimbing dislocations will be dragged along and this will cause the syringe effect---that is, the superflow in the directions $\pm $Y.
Alternatively, in the inverse version, injecting matter into the solid along Y-direction should cause the shear strain with the components $u_{xz}$. [The experimental detection might be easier to carry out in such a setting  (see Ref.~\cite{Kuklov2019})].

Another option  consists of imposing stress $\sigma^{(e)}_{zz}$ which should produce the strain with the components $u_{xz}$.
In a perfect $\it hcp$ crystal such a dependence is forbidden by the symmetry (see, e.g., \cite{Landau}). It, however, can be induced by the superclimbing dislocations bound to the basal ones. Indeed, the stress $\sigma^{(e)}_{zz}$ produces force on the superclimbing dislocation which causes its climb. Accordingly, motion of these dislocations (supported by the superflow along $\pm$Y directions) will be accompanied by glide of the basal dislocations along $\pm $X directions.

It is, however, important to realize that the sign of the effect in each setting described above depends on the product of the ``charges" of the superclimbing and 
basal dislocations---that is, the orientations of their Burgers vectors. 
Thus, on average in a sample containing no excess of basal and superclimbing dislocations with definite ``charges" no effect will occur. 
This, of course, does not exclude statistical fluctuations from sample to sample (see Ref.~\cite{Kuklov2019}). Inducing preferential deformations of a sample with definite signs of the Burgers vectors will determine a specific value of the average (over samples) outcome in each setting.  
Single \he4 crystals are expected to have the strongest signal by avoiding sign-averaging in polycrystalline samples.

\subsection{Dynamic macroscopic pathways}

It is conceivable that the conducting pathways do not exist in equilibrium but are instead established in response to the 
external bias leading to the injection of \he4 atoms from Vycor rods (or from one part of the solid to another) in the from of 
prismatic loops. As discussed in Ref.~\cite{Kuklov2015}, such loops become unstable against inflation (and shape proliferation) 
for large enough chemical potential bias. The threshold for the inflation is inversely proportional  
to the length $R$ of a free segment of a superclimbing dislocation which transforms into the loop under the bias. As discussed above, this mechanism is responsible for growing the solid at pressures above the melting line \cite{Hallock, Hallock2010}. Injected loops may grow macroscopically large and establish the superflow pathways through the whole sample.
This option has been suggested in Ref.~\cite{Kuklov2015}.  Detecting the threshold for inducing the superflow and the syringe effects will be a strong indication for such a mechanism.  
It has also been suggested that in short samples ($8\, \mu$m long) the flow is supported by straight screw dislocations connecting both Vycor terminals \cite{Moses2019}. As argued in Ref.~\cite{Kuklov2015}, screw dislocations can develop helical instability which should also lead to the syringe effect. While there should be no threshold for the superflow in this scenario, there should be a threshold for the chemical potential bias inducing the syringe response.
It can be estimated as $\approx 3-10$mbar.
  
\subsection{Dependence of the superflow on distance between the Vycor ``electrodes"}
As found in Ref.~\cite{Moses2019}, the flow rate decreases logarithmically with the distance between the rods.   This conclusion is based on the data obtained from different samples grown in cells with different geometry. The nature of this dependence is not clear.
Thus, it is important to verify and further quantify this result by designing a long cell with several Vycor rods set along its length.

\subsection{Plasticity induced superflow pathways}
It is important to understand the relationship between the plasticity and superflow in solid \he4. Among the underlying options is the possibility that plastic deformation induces the conducting pathways.  These could be made of dislocations with superfluid cores \cite{screw,sclimb}, superfluid grain boundaries \cite{GB}, and even
superglass regions \cite{SG}. The experiments should be based on combining the techniques \cite{Hallock} and \cite{Beamish2018} in order to carry on simultaneous measurements of the plastic and superflow responses. 

In {\it hcp} \he4 a planar plastic deformation along the slip plane (basal plane) (as utilized in Refs.~\cite{Beamish2018,Zhuch}) should mostly result in creating basal dislocations rather than superclimbing ones. This, however, does not exclude that pile up of the basal dislocations induces conducting channels without any superclimbing dislocations--once the resulting strain in the pile up exceeds 10-15\% (see in Ref.~\cite{stress}) such channels where the vacancy gap vanishes should open up. The channels should exist along the lines where the stress has the square root singularity (see Ref.~\cite{Landau}). As elaborated below, these channels may contain the intriguing phase of solid \he4---the so called superglass \cite{SG}.   

\subsection{Superglass}

One should clearly distinguish intrinsic and extrinsic types of disorder: while the former is believed to induce superfluid properties 
in \he4 crystals, the latter may work in the opposite direction. When \he4 solid was grown inside the silica aerogel \cite{Moses2019}
no flow through the sample has been detected. One plausible explanation is that silica particles disrupt the 
percolating network by providing numerous termination points for superfluid cores.

 The amount of intrinsic disorder can be increased by ``cold working"---applying drastic mechanical deformations to \he4 samples at temperatures below 0.5K (above which  vacancy diffusion becomes effective \cite{Beamish2018}) to produce strains well above $\sim 0.4$\%, as used in Ref.~\cite{Beamish2018}.
There were several attempts to induce disorder by non-planar objects  (see the review \cite{Beamish2019}) starting from the experiment \cite{Andreev2}. If experiments can succeed in increasing the dislocation density to the point when the notion of the regular solid
is lost at the microscopic scale, the possibility of an intriguing metastable superglass phase \cite{SG} opens up. Superglass is 
characterized by a unique combination of two properties: it has finite superfluid density and inhomogeneous disordered average 
density profile. This option is of essentially quantum origin; it should be contrasted with all-classical model of plasticity and work hardening of solid \he4 developed in Refs.~\cite{Balatsky}.

In classical materials, dislocation pile ups lead to fractures. In contrast, in \he4 fractures are not possible because the crystal
will instead melt and try to re-crystallize to fill the void. If the re-crystallization process is dynamically arrested by strong gradients of induced stress, an amorphous
solid with superfluid properties, or superglass, may form.   This should result in the hardening of the material and enhanced supertransport,
as predicted in Ref.~\cite{SG}.    

Another option that appears plausible to us is a shock-type dramatic increase of pressure in low-temperature superfluid \he4. We are talking of the process that takes the density of the liquid not only well above the solidification point, but also well above the metastability region. Since the entropy of the initial state can be arbitrarily low, this opens the door for a peculiar scenario of quantum jamming,
when the shock compression directly converts the liquid into a low-temperature metastable state of superglass.

 \section{Discussion}
The nature of the conducting pathways responsible for the superflow through the solid and the syringe effect in the experiments \cite{Hallock,Moses,Beamish} is one of the most enigmatic problems in the field of strongly interacting many body systems. The current view is that a network of dislocations with  superfluid core is responsible for the effects. While there is a natural qualitative explanation for the syringe effect in terms of the superclimb \cite{sclimb}, quantitative explanation of the observed properties is still lacking.

Here a general assessment of the role of the superclimb indicates that a stable multi connected network can be formed because of binding between superclimbing and basal dislocations which form their own stable network of Y-type junctions. Alternatively,  the pathways should be independent dislocations either static or formed dynamically under the external bias by chemical potential. The proposed experiments should clarify the situation. 
Creating conditions of strong disorder and observing how the response changes (if any) is also an important part for future development of the research in the field of solid \he4.

From the perspective of theory, dynamical properties of the flow pose a very intriguing challenge in the context of the dependence (\ref{F}).  
As discussed in Ref.~\cite{sclimb}, one long superclimbing dislocation in ideal crystal represents a non-Luttinger liquid because its spectrum is quadratic. In a network, this spectrum crosses over to the linear one at momenta smaller than a typical distance $R$ between the pinning points. In samples with $R\gg a$, the compressibility of such a dislocation scales as $\sim R^2$. Accordingly, the Luttinger parameter acquires the factor $ \sim R/a \gg1$, which, on one hand, implies that phase slips should be irrelevant. On the other hand, the dislocation shape  becomes quite fragile because even a small external bias $\sim 1/R$ can lead to the instability of the Bardeen-Herring type resulting in creation and disconnection of prismatic loops of size $\sim R$ from the main dislocation \cite{Kuklov2015}. This mechanism produces giant phase slip events. 

 Along with the challenge of understanding the structure of pathways responsible for the supertransport in currents experimental setups, there is a big question of possibility to control the superfluid disorder by special protocols of sample preparation, including an option of creating the superglass phase. Cold-working techniques seem to be a natural way of exploring this intriguing area. The shock-type compression of the low-temperature superfluid well above the solidification point is yet another intriguing option to explore. 

\begin{acknowledgments}

We thank Robert Hallock,  Moses Chan and Sebastien Balibar  for useful discussions.
This work was supported by the National Science Foundation under the grants DMR-1720251 and DMR-1720465.

\end{acknowledgments}

\end{document}